**Title: Personalized Multimorbidity Management for Patients with Type 2 Diabetes Using Reinforcement Learning of Electronic Health Records**

**Running Title: Multimorbidity Management Using Reinforcement Learning**

Hua Zheng[1], Ilya O. Ryzhov[2], Wei Xie[1,*], and Judy Zhong[3,*]

Affiliations

1. Department of Mechanical and Industrial Engineering, Northeastern University, 360 Huntington Avenue, Boston, MA 02115
2. Robert H. Smith School of Business, University of Maryland, 4347 Van Munching Hall, College Park, MD 20742
3. Division of Biostatistics, Department of Population Health, New York University School of Medicine, 180 Madison Avenue, New York, NY 10016

**Corresponding author[*]:**

Judy Zhong
Associate Professor and Director
Division of Biostatistics Department of Population Health
NYU Langone Health
180 Madison Avenue, 4th Floor, Room 452 New York, NY 10016
Tel: 6465013646
judy.zhong@nyulangone.org

Wei Xie
Assistant Professor
Mechanical and Industrial Engineering, Northeastern University
360 Huntington Avenue, 334 SN, Boston, MA 02115,
Tel: 6173732740
w.xie@northeastern.edu

Word count: 3,213

Number of tables: 7 tables

Number of figures: 3 figures

**ABSTRACT**

**AIMS:** Comorbid chronic conditions are common among people with type 2 diabetes. We developed an Artificial Intelligence algorithm, based on Reinforcement Learning (RL), for personalized diabetes and multimorbidity management with strong potential to improve health outcomes relative to current clinical practice.

**METHODS:** We modeled glycemia, blood pressure and cardiovascular disease (CVD) risk as health outcomes using a retrospective cohort of 16,665 patients with type 2 diabetes from New York University Langone Health ambulatory care electronic health records in 2009 to 2017. We trained a RL prescription algorithm that recommends a treatment regimen optimizing patients' cumulative health outcomes using their individual characteristics and medical history at each encounter. The RL recommendations were evaluated on an independent subset of patients.

**RESULTS:** The single-outcome optimization RL algorithms, RL-glycemia, RL-blood pressure, and RL-CVD, recommended consistent prescriptions with what observed by clinicians in 86.1%, 82.9% and 98.4% of the encounters. For patient encounters in which the RL recommendations differed from the clinician prescriptions, significantly fewer encounters showed uncontrolled glycemia (A1c>8% on 35% of encounters), uncontrolled hypertension (blood pressure > 140mmHg on 16% of encounters) and high CVD risk (risk > 20% on 25% of encounters) under RL algorithms than those observed under clinicians (43%, 27% and 31% of encounters respectively; all $P < 0.001$).

**CONCLUSIONS:** A personalized reinforcement learning prescriptive framework for type 2 diabetes yielded high concordance with clinicians' prescriptions and substantial improvements in glycemia, blood pressure, cardiovascular disease risk outcomes.

**Key Points**

- Artificial intelligence (AI) prescription algorithms have been successfully applied to single disease problems, but previous applications have not considered comorbid conditions, pharmacological treatments, treatment histories, and other individual characteristics that are important for personalized diabetes management.

- We trained and evaluated a series of AI algorithms to optimize patients' glycemia, blood pressure and CVD risk outcomes, either individually or jointly, using a retrospective cohort of T2DM patients from an ambulatory care electronic health records database (2009-2017).

- When optimizing glycemia, blood pressure and CVD risk individually, the algorithms recommended prescriptions consistently with clinicians' decisions in 86.1%, 82.9% and 98.4% of patient encounters. In cases where the AI recommendation differed from the clinician's prescription, health outcomes were significantly improved.

- The RL algorithm can be integrated into EHR platforms to assist physicians with dynamic real-time suggestions on personalized treatment paths.

# 1. INTRODUCTION

Comorbid chronic conditions are common among people with Type 2 Diabetes (T2DM) (1). Hypertension (HTN) and atherosclerotic cardiovascular disease (CVD) are the two most common multimorbidities for T2DM patients (2). Therefore, the need to address comorbid chronic conditions, in addition to patients' diabetes-specific treatment goals (3) poses a substantial challenge for effective T2DM management. Although improvements in glycemic monitoring and control have been documented in several large systems of care, and more widespread use of treatments such as angiotensin-converting enzyme (ACE) inhibitors and aspirin have decreased patients' risk of cardiovascular death, the current commonly used standard of care and guidelines are usually built around single diseases (4). Despite the increasing numbers of patients with multimorbidity, such patients are usually excluded from randomized controlled trials (5-7). A systematic review of managing patients with multimorbidity identified only 10 randomized trials worldwide and highlighted the paucity of research into interventions to improve outcomes for patients with multimorbidity (8). On the other hand, there is a large volume of evidence suggesting that the response to T2DM treatment, HTN treatment and CVD prevention differs between population subgroups (9; 10). Therefore, the need for an individualized approach is especially pressing given the variety of comorbid conditions, pharmacological treatments, individual treatment histories, and other individual characteristics that may inform treatment selection.

We provide an artificial intelligence (AI) prescription algorithm, based on reinforcement learning (RL), which is able to dynamically suggest personalized optimal treatments for patients with T2DM to manage their multimorbidity based on evidence from patients' electronic health records (EHRs). Reinforcement learning has been successfully applied in the past to single disease problems, such as blood glucose control (11), HIV therapy (12), cancer treatment (13), anemia treatment in hemodialysis patients (14), treatment strategies for sepsis in intensive care (15) and personalized regime of sedation dosage and ventilator support for patients in Intensive Care Units (ICUs) (12). Prescriptive algorithms using regression trees and $k$ nearest neighbors (kNN) have previously shown great potential in personalized diabetes management (16; 17).

Our approach leverages the power of RL and abundant data in the EHR system to dynamically recommend treatment prescriptions, which are personalized based on patient characteristics, including age, sex, race, BMI, blood pressure (BP), lab tests, duration of T2DM and treatment history. In our setting, we first apply RL to optimize glycemic control, BP control, and CVD prevention separately, and then study the potential of RL for multimorbidity management by optimizing all three outcomes jointly. We evaluate the effectiveness of the personalized treatment recommendations made by RL against the observed clinician's treatment by estimating patients' outcomes based on the outcomes of similar patients in the EHR database.

## 2. RESEARCH DESIGN AND METHODS

### 2.1 Study Design and Participants

We used ambulatory care EHR samples for T2DM patients from New York University Langone Health (NYULH-EHR) to derive and validate the RL algorithm. Eligible patients have had at least one encounter with an NYULH ambulatory primary care physician between 2009-2017 and have been selected by a T2DM rule-based phenotyping algorithm, defined as the following criteria: (1) having at least two encounters with an International Classification of Diseases (ICD)-10 code for T2DM, or (2) having ≥ two abnormal hemoglobin A1c (A1c) (≥ 6.5%) and at least one encounter with an ICD-10 code for T2DM, or (3) having a prescription for a T2DM medication, excluding metformin and acarbose. We excluded patients seen for consultation only and patients in emergency department, inpatient or specialist settings, as these lacked consistent documentation of T2DM across encounters. We randomly selected 60% of the eligible patients as the training cohort to develop the RL algorithm and reserved the remaining 40% patients as the test cohort to evaluate the performance of the RL algorithm. This study was approved by the NYULH IRB and the data were de-identified to ensure anonymity.

For each patient, we had access to demographic data, including age, sex, race, ethnicity and smoking status, as well as the following biomarkers: systolic BP (SBP), diastolic BP (DBP), body mass index (BMI), HbA1c, total cholesterol (TC), low-density lipoprotein (LDL), high-density lipoprotein (HDL), creatinine, triglycerides and estimated glomerular filtration rate (eGFR). In NYULH-EHR, 1% of samples had missing vitals including blood pressures and BMI, 8% missing HbA1c, 5-32% missing renal function biomarkers, and 13% missing lipid biomarkers. Following Lundberg et al.(18), we imputed the missing patients' biomarkers based on the observed values measured in previous encounters.

Medication prescriptions were first grouped by therapeutic class codes of antihyperglycemic, antihypertensive and lipid-lowering, then analyzed by pharmacologic subclass. The antihyperglycemic therapeutic class contains nine pharmacologic subclasses including PPARs agonist thiazolidinedione (PPARg), insulin release stimulant type (INSR), incretin mimetic (GLP-1 receptor agonist) (GLP1), DPP-4 inhibitor and biguanide (DPP4-BIG), DPP-4 inhibitors (DPP4), biguanide type (BIG), insulin release stimulant and biguanide (INSR-BIG), sodium-glucose cotransport-2 inhibitors (SGLT2) and insulins (INSO). The antihypertensive therapeutic class contains ten pharmacologic subclasses including angiotensin receptor antagonists (ARA), Potassium-sparing diuretics in combination (PSD), alpha/beta-adrenergic blocking agents (ABAB), ACE inhibitor with thiazide or thiazide-like diuretic (ACE-TD), angiotensin receptor antagonists with thiazide diuretic (ARA-TD), ACE inhibitors (ACE), thiazide and related diuretics (TD), beta-adrenergic blocking agents (BAB), calcium channel blocking agents (CCB) and angiotensin receptor antagonists with calcium channel blocking agents (ARA-CCB). The antihyperlipidemic therapeutic class contains five pharmacologic subclasses including bile salt sequestrants (BSS), HMG-CoA reductase inhibitors (HMG), HMG-CoA reductase inhibitors and cholesterol absorption inhibitors (HMG-CA), proprotein convertase subtilisin/kexin type 9 inhibitors (PCSK9) and lipotropics (LIP).

## 2.2 Overview of RL algorithm

RL algorithms model the course of patients' EHR histories, which includes prescriptions, biomarkers and health outcomes changing over time using a Markov decision process with key elements including state, action, and reward (15; 19). In this setting, "state" refers to the observed patient demographics, laboratory test results at the current encounter and their histories of lab tests and prescriptions. "Action" refers to the prescribed treatment regimen at the current encounter, which are pharmacologic subclasses or their combinations. The result of an action is a numerical reward representing the improvement of health outcomes compared to the previous encounter. The cumulative reward is defined as the sum of the rewards along the course of EHR encounter records. RL has been well established as an efficient AI learning algorithm to maximize cumulative reward by selecting an optimal action at each encounter through a learning algorithm called Deep Q Networks (20; 21) with a multi-layer (deep) neural network. An important advantage of RL is that the action in every encounter is personalized to the patient's individual characteristics as they are observed, in a way that optimizes the cumulative reward. In this paper, we focus on glycemia (lowering A1c towards 6.5%) control, BP (lowering SBP towards 120mmHg) control, and CVD (minimizing CVD risk) prevention. We first optimize each outcome individually using three separate RL algorithms, referred to as RL-glycemia, RL-BP, and RL-CVD. We then train a multimorbidity management RL algorithm (RL-multimorbidity) to optimize glycemia, BP and CVD risk simultaneously. The details of state, action and reward are described as following:

- **State**: a list of observed patient characteristics including age, sex, race, smoking status; vitals and lab test values at current encounter and in the past 6 months including BMI, weight, SBP, DBP, triglycerides, TC, HDL, LDL, A1c and creatinine; prescription history in the past 6 months; and encounter histories including days since the previous encounter, and days since the first encounter.
- **Action**: The action space consists of the pharmacologic subclasses and their combinations, referred to as the treatment regimen. The action space of RL-glycemia contains 9 pharmacologic subclasses in the antihyperglycemic therapeutic class or their combinations. The action space of RL-BP contains 10 pharmacologic subclasses in the antihypertensive therapeutic class or their combinations. The action space of RL-CVD contains 5 pharmacologic subclasses in the antihyperlipidemic therapeutic class or their combinations. The action space of RL-multimorbidity contains pharmacologic subclasses in all three therapeutic classes or their combinations.

- **Reward:** The reward of a prescription is a numeric measure of treatment efficacies between two consecutive encounters. For RL-glycemia, if A1c<5.6% in both encounters, their rewards are zero, otherwise the reward is defined by the reduction in A1c. For RL-BP, if patients have no HTN symptom (<120 mmHg) in both encounters, the reward is zero, otherwise it is equal to the decrease in SBP. For RL-CVD, the reward is the reduction in global CVD Framingham Risk Score (FRS) (22), which is a function of age, TC, HDL, SBP, treatment for hypertension, smoking, and T2DM status (all yes). Sex-specific risk equations were applied to males and females separately. For RL-multimorbidity, the reward is defined as the average of standardized rewards values of RL-BP, RL-glycemia, and RL-CVD (model and training details in Supplementary Materials).

**2.3 Model Evaluation**

We evaluated the RL-recommended therapy by comparing its effect with the observed clinician's prescriptions on the test cohort of NYULH EHR samples. In each encounter, the RL algorithm recommends a treatment regimen for the patient. If the recommendation is the same as the observed clinician's prescription in the data, we say that RL is "consistent" with the clinicians. When RL is discrepant with the clinician's prescription, the efficacy of the RL-recommended treatment is not directly observed. For this reason, we impute the outcome of the RL-recommended treatment using kNN regression, an approach commonly used for causal inference in observational studies (23). In short, the imputation works by averaging the outcomes of the $k$ most similar patient encounters, in terms of patient characteristics, in which the RL-recommended therapy had been administered by clinicians. The similarity between patient encounters was estimated by Euclidean distance as in Bertsimas et al. (16). To assess the performance of the imputation, we first compared imputed outcomes with observed outcomes under clinician's treatments, and found 87-95% correlation between them, indicating that the imputation algorithm can effectively estimate unobserved health outcomes (Table 1). We varied the number *k* of nearest neighbors and found the performance of the imputation (for any of the three health outcomes) was insensitive when *k* was between 8 and 10. We estimated the efficacy of the recommendations made by RL first in the whole set of test samples, then for individual gender, racial and age subgroups.

**Table 1 Counterfactual outcome versus true clinical outcome comparison based on kNN regression.**

| Biomarkers | Counterfactual Outcome | True Outcome | Pearson Correlation |
|---|---|---|---|
| BP Systolic | 128.55 (0.017) | 128.68 (0.022) | 0.89 |
| BP Diastolic | 74.27 (0.010) | 74.29 (0.014) | 0.89 |
| Triglycerides | 152.85 (0.118) | 153.74 (0.14) | 0.87 |
| Total Cholesterol | 174.05 (0.056) | 174.31 (0.061) | 0.93 |
| HDL Cholesterol | 51.37 (0.022) | 50.37 (0.024) | 0.95 |
| LDL Cholesterol | 92.73 (0.047) | 92.85 (0.052) | 0.92 |
| A1c | 7.02 (0.002) | 7.05 (0.002) | 0.92 |

**2.4 Feature Importance**

To better understand which features have the most impact on treatment recommendations, we used SHAP (SHapley Additive exPlanations) (24; 25) to estimate and rank the contributions of clinician features explaining RL and clinician prescriptions.

## 3. RESULTS

Overall, 16,665 patients in NYULH ambulatory care EHR samples had a query based T2DM diagnosis in 2009 to 2017, with 1,278,785 encounters (median 12 encounters per patient). The number of T2DM patients was robust to variations in the T2DM phenotyping algorithm resulting from changes in the required number of encounters with T2DM ICD-10s and the medications. The demographic and clinic characteristics of the analysis cohort are shown in Table 2. Overall, the patients were 65.6 years old, comprised of 8,278 females (54.6%). On average, T2DM patients showed A1c 7.1% and SBP 128.9 mmHg. Antihyperglycemic, antihypertensive medications and antihyperlipidemic medications were prescribed in 665,768 (52.1%), 849,328 (66.4%) and 428,427 (33.5%) encounters respectively. The median follow-up time was 2.6 years since T2DM diagnosis (interquartile range [IQR]: 1.9-3.9 years). We first trained the RL algorithms using 530,786 (60%) T2DM patient encounters, and then assessed their performance using the remaining 394,447 (40%) T2DM patient encounters.

**Table 2 Demographics and clinic characteristics of NYULH-EHR patients with type 2 diabetes.**

| Demographics and clinic characteristics | Number of Patients (N=16,665) |
|---|---|
| Age (years, Mean (SD)) | 65.62 (13.66) |
| Male (N (%)) | 6876 (45.37) |
| Race (N(%)) | |
| African American | 5,146 (33.96) |
| native American | 55 (0.36) |
| Asian | 692 (4.57) |
| Caucasian (White) | 7,888 (52.05) |
| smoker (ever and current, (N%)) | 1,043 (6.88) |
| Systolic Blood Pressure (mmHg, Mean (SD)) | 128.93 (14.60) |
| Diastolic Blood Pressure (mmHg, Mean (SD)) | 74.19 (8.88) |
| Body Mass Index (kg/m$^2$, Mean (SD) | 31.56 (6.86) |
| Triglycerides (mg/dL, Mean (SD)) | 155.06 (91.97) |
| Creatinine (mg/dL, Mean (SD)) | 1.02 (0.44) |
| Total Cholesterol (mg/dL, Mean (SD)) | 173.37 (39.82) |
| Low-density Lipoproteins (mg/dL, Mean (SD)) | 91.99 (33.53) |
| High-density Lipoproteins (mg/dL, Mean (SD)) | 51.00 (15.25) |
| A1c (%, Mean (SD)) | 7.11 (1.46) |
| Medications | Number of Patient Encounters (n=1,278,785) |
| Antihyperglycemic Class (N(%)) | 665,768 |
|   biguanide type (BIG) | 250,438 (37.62) |
|   insulin release stimulant type (INSR) | 110,139 (16.54) |
|   insulins (INSO) | 106,356 (15.97) |
|   DPP-4 inhibitors (DPP4) | 64,090 (9.63) |
|   DPP-4 inhibitor and biguanide (DPP4-BIG) | 53,337 (8.01) |
|   incretin mimetic (GLP-1 receptor agonist) (GLP1) | 35,696 (5.36) |
|   sodium-glucose cotransport-2 inhibitors (SGLT2) | 23,021 (3.46) |
|   PPARs agonist thiazolidinedione (PPARg) | 12,573 (1.89) |
|   insulin release stimulant and biguanide (INSR-BIG) | 10,118 (1.52) |
| Antihypertensive Class (N(%)) | 849,328 |
|   beta-adrenergic blocking agents (BAB) | 200,114 (23.56) |
|   calcium channel blocking agents (CCB) | 151,701 (17.86) |
|   ACE inhibitors (ACE) | 149,561 (17.61) |
|   angiotensin receptor antagonists (ARA) | 138,705 (16.33) |
|   angiotensin receptor antagonists with thiazide diuretic (ARA-TD) | 67,964 (8.00) |
|   alpha/beta-adrenergic blocking agents (ABAB) | 57,426 (6.76) |
|   thiazide and related diuretics (TD) | 57,196 (6.73) |
|   ACE inhibitor with thiazide or thiazide-like diuretic (ACE-TD) | 14,486 (1.71) |
|   potassium-sparing diuretics in combination (PSD) | 6,246 (0.74) |
|   angiotensin receptor antagonists with calcium channel blocking agents (ARA-CCB) | 5,929 (0.70) |
| lipid-lowering Class (N(%)) | 428,427 |
|   HMG-CoA reductase inhibitors (HMG) | 379,924 (88.68) |
|   lipotropics (LIP) | 40,173 (9.38) |
|   bile salt sequestrants (BSS) | 5,262 (1.23) |
| HMG-CoA reductase inhibitors and cholesterol absorption inhibitors (HMG-CA) | 2,286 (0.53) |
| proprotein convertase subtilisin/kexin type 9 inhibitors (PCSK9) | 782 (0.18) |

Categorical variables are summarized with frequencies (percentages) unless otherwise indicated. Continuous variables are summarized as the mean (standard deviation) of biomarkers.

**Table 3 Performance of RL algorithms with comparison between RL and clinicians for glycemic control, hypertension control, and CVD prevention.**

| | | | |
|---|---|---|---|
| **RL-glycemia** | | | |
| Encounters for which algorithm's recommendation differed from observed Clinician's prescription (N(%)) | 15,578 (13.9) | | |
| | RL-glycemia | Clinician's prescription | P-value |
| A1c (Mean(SE)) | 7.80 (0.01) | 8.09 (0.01) | <0.001 |
| A1c > 8% (N(%)) | 5,421 (34.8) | 6,617 (42.5) | <0.001 |
| **RL-BP** | | | |
| Encounters for which algorithm's recommendation differed from observed Clinician's prescription (N(%)) | 20,251 (17.1) | | |
| | RL-BP | Clinician's prescription | P-value |
| SBP(Mean(SE)) | 131.77(0.06) | 132.35 (0.11) | <0.001 |
| SBP > 140 mmHg (N(%)) | 3,256 (16.1) | 5,390 (26.6) | <0.001 |
| **RL-CVD** | | | |
| Encounters for which algorithm's recommendation differed from observed Clinician's prescription (N(%)) | 946 (1.6) | | |
| | RL-CVD | Clinician's prescription | P-value |
| FHS (Mean(SE)) | 13.65 (0.26) | 17.18 (0.36) | <0.001 |
| FHS > 20% (N(%)) | 237 (25.1) | 299 (31.6) | <0.001 |
| **RL- multimorbidity** | | | |
| Encounters for which algorithm's recommendation differed from observed Clinician's prescription (N(%)) | 102,184 (28.9) | | |
| | RL-multimorbidity | Clinician's prescription | P-value |
| A1c (Mean(SE)) | 7.14 (0.003) | 7.19 (0.005) | <0.001 |
| A1c > 8% (N(%)) | 16,436 (16.08) | 20,879 (20.43) | <0.001 |
| SBP (Mean(SE)) | 129.40 (0.03) | 129.58 (0.05) | <0.001 |
| SBP > 140 mmHg (N(%)) | 9,800 (9.59) | 20,957 (20.51) | <0.001 |
| FHS (Mean(SE)) | 21.89 (0.04) | 25.61 (0.05) | <0.001 |
| FHS > 20% (N(%)) | 48,283 (47.3) | 55,957 (54.8) | <0.001 |

The performance of the RL algorithms on the test dataset is summarized in Table 3. The RL-glycemia algorithm was consistent with clinicians' prescriptions in 86.1% of encounters. In the remaining 15,578 (13.9%) encounters, the mean A1c under clinician-prescription was 8.09% (95% CI: 8.06-8.12), while the mean A1c under RL-glycemia was 7.80% (95% CI: 7.78-7.82), showing a 0.30% (95% CI: 0.28-0.32) reduction (P<0.001). Significantly fewer encounters showed uncontrolled A1c (A1c>8%) under RL-glycemia than under clinicians (35% vs 43%, P < 0.001). The RL-BP algorithm was consistent with clinicians' prescriptions in 82.9% of encounters. In the remaining 20,251 encounters (17.1%) with discrepant recommendations, RL-BP achieved a 0.58 mmHg (95% CI: 0.37, 0.79) reduction in SBP relative to clinicians' prescriptions (131.77 vs 132.35 mmHg, P<0.001). Fewer encounters showed uncontrolled HTN (SBP>140mmHg) under RL-BP than under clinicians' prescriptions (16% vs 27%, P < 0.001). The RL-CVD was consistent with clinicians' prescriptions in 98.4% of encounters. In the remaining 946 encounters (1.6%) with discrepant recommendations from RL and clinicians', the mean FRS reduced 3.53% (95% CI: 2.94, 4.12) under RL-CVD from under clinician-prescription (13.65% vs 17.18%, P<0.001), with fewer encounters showing high FRS risk (>20%) (25% vs 31%, P < 0.01). These results collectively showed high concordance between the optimized RL algorithms and clinicians' prescriptions for single target management for patients with T2DM. However, there were more frequent discrepancies between RL-multimorbidity and clinicians. The RL-multimorbidity algorithm was consistent with clinicians' prescriptions in 71.1% of encounters. In the remaining 102,184 encounters (28.9%) with discrepant prescriptions, 16,436 (16.1%), 9,800 (9.6%) and 48,283 (47.3%) encounters had uncontrolled A1c, uncontrolled HTN and high FRS risk that were significantly lower than observed outcomes under clinician's prescriptions (20.4%, 20.5% and 54.8% respectively).

To understand when and how RL makes different prescriptions from clinicians, Table 4 compares consistent and discrepant encounters by patient demographics and clinic characteristics. The most significantly associated factor was the severity at the time of the encounter. For RL-glycemia, encounters with higher A1c were more likely to have different recommendations (average A1c 8.1% for discrepant encounters vs 7.5% for consistent encounters, P<0.001). For RL-BP, encounters with higher SBP were more likely to have different recommendations (average SBP 132.85 vs 131.00 mmHg, P<0.001).

The efficacy of the RL prescriptive algorithms was consistently observed across T2DM patient, gender, racial and age subgroups (Table 5-7). Specifically, African American (AA) T2DM patients, and T2DM patients of age older than 60 observed higher efficacies from the RL algorithms than clinicians' prescriptions as compared to the observed efficacies in white patients and in patients of age 60 and younger. For example, A1c under RL-glycemia for AA patients were 0.39% lower than under clinician's treatment. In contrast, A1c under RL-glycemia were 0.28% lower than that under clinician's treatment for white patients. Patients of age 60 and younger observed higher efficacy, with A1c under RL-glycemia 0.47% lower than that under clinician's treatment, than those older than 60 with A1c under RL-glycemia 0.19% lower than that under clinician's treatment.

**Table 4 Comparison of RL and clinicians for glycemic control, BP control and CVD prevention. Demographic characteristics of patients having encounters at which RL and clinicians prescribed consistently versus differently.**

| Features | T2DM (n=15578) | | HTN (n=20251) | | CVD (n=946) | | Multimorbidity (n=102184) | |
|---|---|---|---|---|---|---|---|---|
| Prescriptions Consistency (%) | No (13.89) | Yes (86.11) | No (17.08) | Yes (82.82) | No (1.63) | Yes (98.37) | No (28.88) | Yes (71.12) |
| Age (years) | 65.89 (13.77) | 64.25 (13.69) | 69.42 (12.54) | 68.79 (12.79) | 68.39 (11.65) | 68.87 (12.10) | 66.24 (13.36) | 65.87 (13.64) |
| Percent male | 47.21 | 45.34 | 43.39 | 43.65 | 54.50 | 46.41 | 45.59 | 44.89 |
| Percent black | 33.53 | 34.61 | 31.15 | 32.45 | 16.27 | 26.85 | 33.40 | 33.68 |
| Percent native American | 0.51 | 0.44 | 0.28 | 0.23 | 0.53 | 0.32 | 0.39 | 0.37 |
| Percent Asian | 4.47 | 4.26 | 4.04 | 3.88 | 3.57 | 4.14 | 4.54 | 4.32 |
| Percent white | 53.91 | 52.07 | 57.80 | 56.62 | 72.88 | 61.49 | 53.30 | 53.18 |
| Percent smoke | 6.98 | 6.82 | 5.64 | 6.01 | 8.99 | 6.14 | 6.89 | 6.67 |
| Systolic BP (SBP) (mmHg) | 127.59 (14.35) | 127.28 (13.73) | 132.85 (16.68) | 131.00 (14.99) | 125.72 (14.08) | 127.52 (13.63) | 131.10 (15.89) | 128.65 (14.30) |
| Diastolic BP (DBP) (mmHg) | 74.38 (8.68) | 74.13 (8.51) | 75.25 (10.12) | 74.36 (9.27) | 73.63 (8.26) | 73.47 (8.28) | 74.13 (9.52) | 74.02 (8.72) |
| BMI (kg/m2) | 31.84 (6.92) | 32.01 (7.19) | 32.33 (6.88) | 31.49 (6.81) | 29.98 (5.69) | 30.76 (6.63) | 32.06 (6.67) | 31.51 (6.94) |
| Triglycerides (mg/dL) | 163.57 (104.98) | 157.51 (96.43) | 155.81 (86.83) | 150.20 (81.06) | 197.54 (163.01) | 159.31 (97.69) | 159.19 (94.16) | 154.27 (89.03) |
| Creatinine (mg/dL) | 0.96 (0.38) | 0.99 (0.41) | 1.05 (0.44) | 1.07 (0.48) | 1.05 (0.43) | 1.02 (0.43) | 1.06 (0.45) | 1.02 (0.45) |
| Total Cholesterol (mg/dL) | 172.25 (39.40) | 172.97 (38.85) | 173.76 (38.87) | 172.81 (38.75) | 180.62 (46.24) | 176.05 (42.67) | 170.21 (39.66) | 173.57 (39.70) |
| LDL Cholesterol (mg/dL) | 90.81 (32.68) | 91.27 (32.80) | 93.00 (33.05) | 92.03 (32.70) | 93.38 (37.30) | 93.46 (35.92) | 89.77 (33.36) | 92.04 (33.45) |
| HDL Cholesterol (mg/dL) | 49.59 (15.07) | 50.89 (15.56) | 50.20 (14.71) | 51.32 (15.31) | 49.77 (16.04) | 51.46 (14.86) | 49.29 (14.34) | 51.29 (15.34) |
| A1c (%) | 8.11 (1.81) | 7.51 (1.62) | 6.95 (1.30) | 6.84 (1.25) | 6.85 (1.29) | 6.82 (1.24) | 7.09 (1.38) | 7.08 (1.43) |

Categorical variables are summarized with frequencies (percentages). Continuous variables are summarized as the mean (standard deviation) of biomarkers.

**Table 5 Subgroup results of glycemic control RL algorithm.**

| Subgroup | Number of encounters | RL benefit relative to clinician policy | | |
|---|---|---|---|---|
| | | A1c under RL | A1c under clinician | Benefit |
| Male | 7072 | 7.87 (0.01) | 8.20 (0.02) | -0.33 (0.02) |
| Female | 8506 | 7.73 (0.01) | 8.00 (0.02) | -0.27 (0.02) |
| Age > 60 | 9548 | 7.63 (0.01) | 7.82 (0.02) | -0.19 (0.01) |
| Age ≤ 60 | 6030 | 8.06 (0.02) | 8.53 (0.03) | -0.47 (0.02) |
| White | 8427 | 7.54 (0.01) | 7.81 (0.02) | -0.28 (0.02) |
| Black | 5181 | 8.16 (0.02) | 8.55 (0.03) | -0.39 (0.02) |
| Other Race | 1970 | 7.94 (0.03) | 8.10 (0.04) | -0.16 (0.04) |
| Smoke | 1026 | 8.08 (0.04) | 8.40 (0.06) | -0.32 (0.05) |
| Non-Smoke | 14552 | 7.78 (0.01) | 8.07 (0.01) | -0.30 (0.01) |

**Table 6 Subgroup results of BP control RL algorithm.**

| Subgroup | Number of encounters | RL benefit relative to clinician policy | | |
|---|---|---|---|---|
| | | SBP under RL | SBP under clinician | Benefit |
| Male | 8108 | 131.32 (0.09) | 132.45 (0.17) | -1.13 (0.17) |
| Female | 12143 | 132.07 (0.08) | 132.29 (0.14) | -0.22 (0.14) |
| Age > 60 | 16151 | 131.43 (0.07) | 132.34 (0.12) | -0.90 (0.12) |
| Age ≤ 60 | 4100 | 133.12 (0.13) | 132.43 (0.25) | 0.68 (0.24) |
| White | 11925 | 130.35 (0.07) | 131.22 (0.14) | -0.87 (0.14) |
| Black | 6536 | 134.19 (0.11) | 135.12 (0.20) | -0.93 (0.19) |
| Other Race | 1790 | 132.41 (0.23) | 129.79 (0.39) | 2.62 (0.38) |
| Smoke | 951 | 132.34 (0.31) | 132.55 (0.54) | -0.21 (0.53) |
| Non-Smoke | 19300 | 131.74 (0.06) | 132.35 (0.11) | -0.60 (0.11) |

**Table 7 Subgroup results of multimorbidity control RL algorithm.**

| Subgroup | Number of encounters | RL benefit relative to clinician policy (standard of care) | | | | | | |
|---|---|---|---|---|---|---|---|---|
| | | A1c | Systolic BP | Triglycerides | Total Cholesterol | LDL Cholesterol | HDL Cholesterol | CVD Risk |
| Male | 43816 | -0.09 (0.01) | -0.32 (0.07) | -5.27 (0.50) | -0.10 (0.17) | 0.08 (0.14) | 0.71 (0.06) | -5.09 (0.09) |
| Female | 58368 | -0.02 (0.01) | -0.07 (0.06) | -1.99 (0.33) | -1.23 (0.16) | -0.61 (0.14) | -0.41 (0.06) | -2.68 (0.05) |
| Age > 60 | 75924 | 0.01 (0.00) | -0.59 (0.05) | -0.43 (0.29) | -0.05 (0.13) | 0.27 (0.12) | -0.24 (0.05) | -5.70 (0.06) |

| | | | | | | | |
|---|---|---|---|---|---|---|---|
| Age ≤ 60 | 26260 | -0.23 (0.01) | 1.02 (0.09) | -11.97 (0.72) | -2.75 (0.25) | -1.99 (0.21) | 0.98 (0.08) | 2.03 (0.07) |
| White | 60029 | -0.02 (0.00) | -0.02 (0.06) | -3.78 (0.37) | -1.60 (0.15) | -1.12 (0.13) | 0.16 (0.06) | -4.04 (0.07) |
| Black | 31775 | -0.12 (0.01) | -0.92 (0.09) | 1.79 (0.47) | -0.52 (0.22) | -0.43 (0.19) | -0.64 (0.08) | -3.39 (0.08) |
| Other Race | 10380 | -0.02 (0.01) | 1.17 (0.15) | -17.02 (1.08) | 3.50 (0.39) | 4.70 (0.33) | 1.78 (0.13) | -2.82 (0.14) |
| Smoke | 5747 | -0.10 (0.02) | -0.52 (0.20) | -16.54 (1.71) | -1.31 (0.55) | -0.73 (0.46) | 2.14 (0.17) | -10.41 (0.26) |
| Non-Smoke | 96437 | -0.05 (0.00) | -0.16 (0.05) | -2.61 (0.28) | -0.71 (0.12) | -0.29 (0.10) | -0.05 (0.05) | -3.31 (0.05) |

**Fig 1**

A

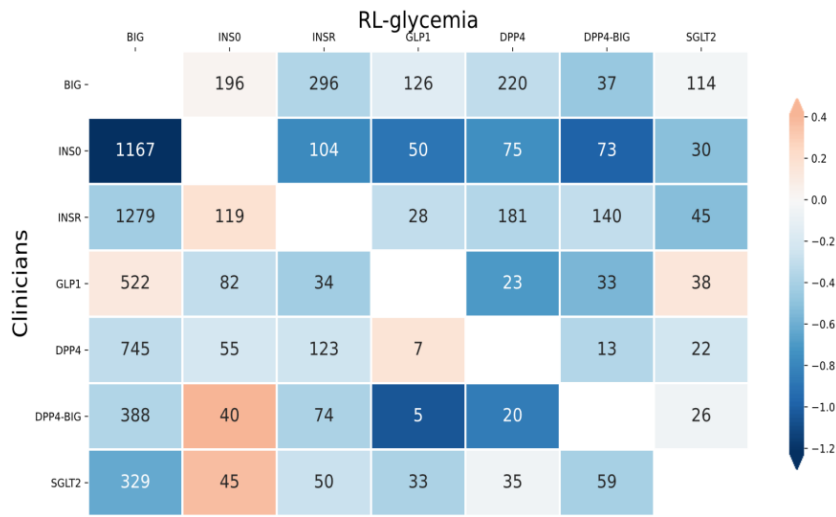

B.

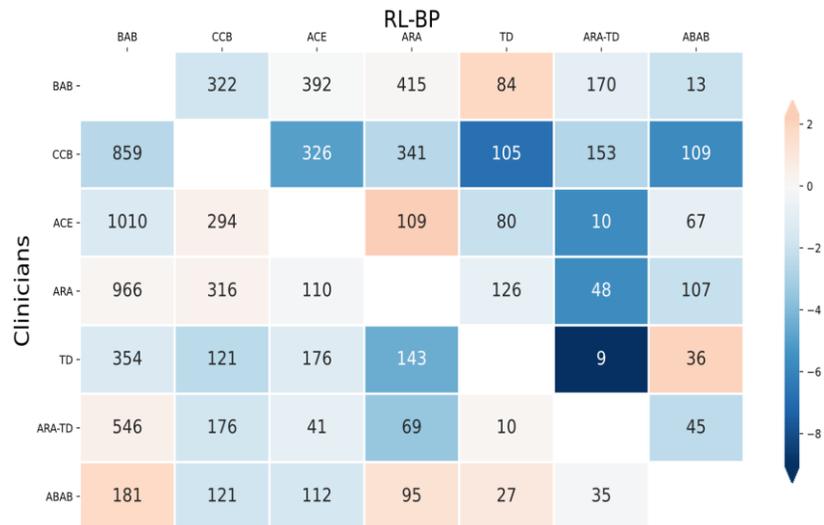

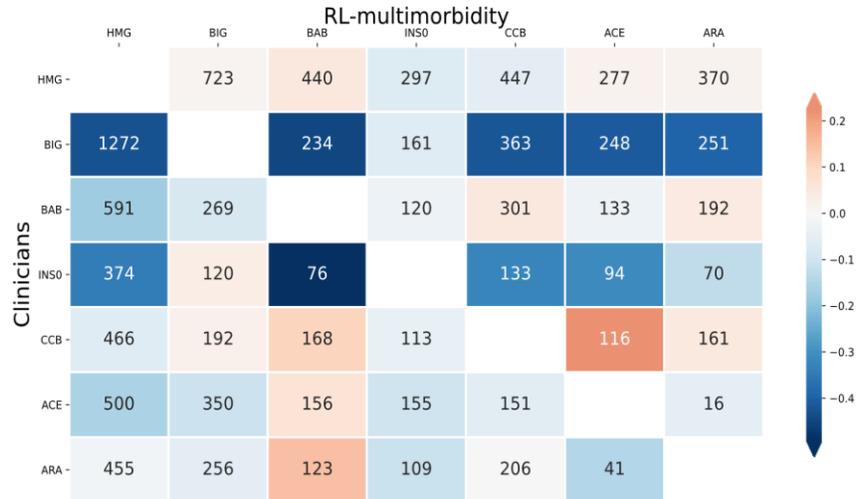

**Patterns of most frequent discrepant RL recommendations and clinicians' prescriptions for RL-glycemia (A), RL-BP (B) and RL-multimorbidity (C).** Each cell and the numbers represent patients for whom RL (labels on horizontal axis) recommended a different regimen from the one given by clinicians (labels on vertical axis). The color in each cell quantifies the improvement in health outcomes achieved by the RL recommendation relative to clinician's prescription, with blue indicating benefits of the RL recommendation and orange indicating worsening outcomes relative to clinician's prescription. (A) indicates the mean A1c reduction (%) of RL-glycemia (labels on horizontal axis) than clinicians (labels on vertical axis); (B) indicates the mean SBP decrease (mmHg) of RL-BP (labels on horizontal axis) than clinicians (labels on vertical axis), and (C) indicates the mean difference of multimorbidity reward from RL-multimorbidity (labels on horizontal axis) than clinicians (labels on vertical axis). RL-CVD was consistent with clinicians' prescriptions for the vast majority of encounters, thus was not shown in this Figure.

The patterns of different treatment recommendations, along with the resulting differences in health outcomes, are illustrated in **Fig. 1** for RL-glycemia, RL-BP and RL-multimorbidity. In the case of RL-glycemia, the most frequently observed discrepancy (1,167 encounters) was that clinicians prescribed insulin monotherapy (INSO) while RL prescribed biguanide type (BIG). On these encounters, RL-glycemia achieved on average 1.22% lower A1c than clinicians. In the case of RL-BP, the most frequently observed discrepancy (1,010 encounters) was that clinicians prescribed ACE inhibitors (ACE) while RL prescribed beta-adrenergic blocking agents (BAB). On these encounters, RL-BP achieved 6.78 mmHg lower SBP. The most frequently observed discrepancy between RL-multimorbidity and clinician's prescription was biguanide type (BIG) prescribed by clinicians and HMG-CoA reductase inhibitors (HMG) prescribed by RL-multimorbidity observed on 1,272 patient encounters. On these discrepant encounters, RL-multimorbidity achieved 0.15% higher A1c, but 2.42% lower CVD risk and 0.30 mmHg lower SBP. Overall, RL algorithms tended to prescribe fewer medications than clinicians (**Fig. 2**).

**Fig 2.**

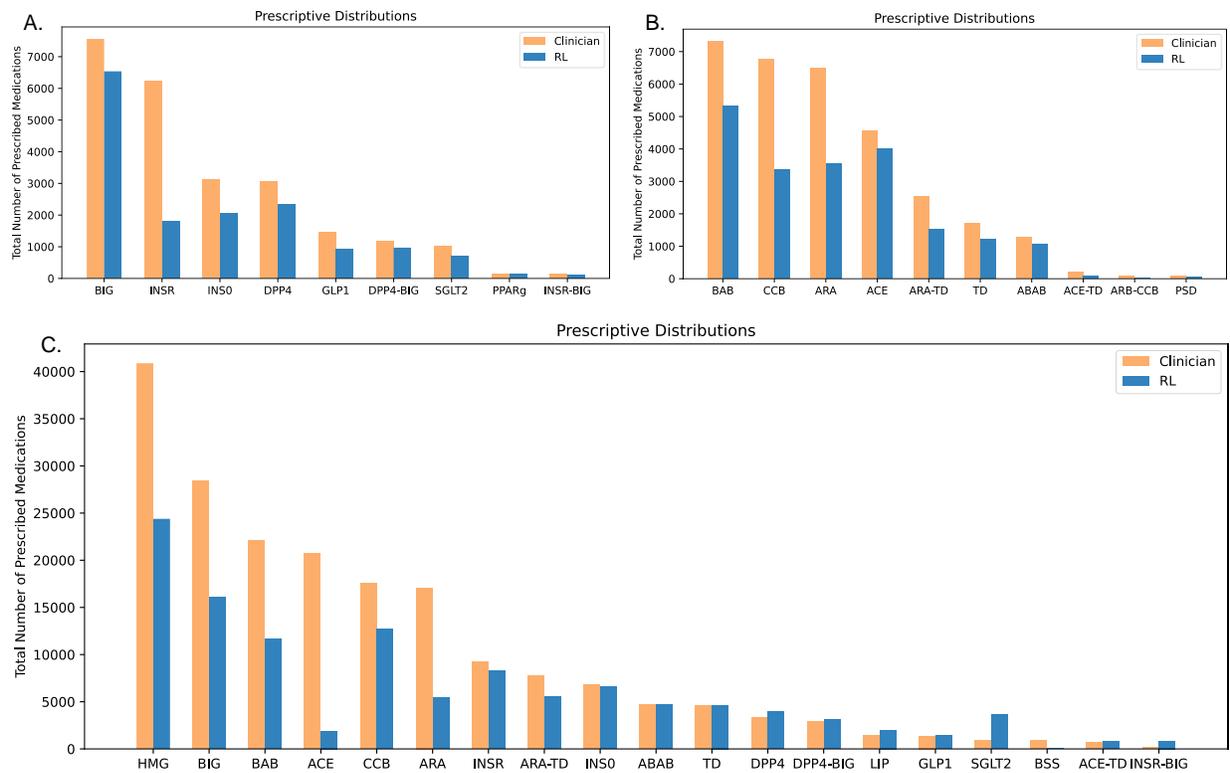

**Prescription medication use by RL versus clinicians. (A) total number of drugs prescribed for blood glucose control; (B) total number of drugs prescribed for BP control; (C) total number of drugs prescribed for multimorbidity management.**

**Fig. 3** shows the importance of features associated with RL-multimorbidity algorithm and clinicians' prescriptions. In general, there was reasonable agreement between the feature-importance estimates of RL-multimorbidity and those identified by the clinicians. A1c is the most important feature for clinicians while RL-multimorbidity was most influenced by recent therapies, age, BMI and A1c. One difference is the importance of creatinine in the clinicians' prescription, while it was not as important for RL-multimorbidity. Another difference is the reduced role of the time since first encounter in RL-multimorbidity as compared to clinicians' prescriptions.

**Fig 3**

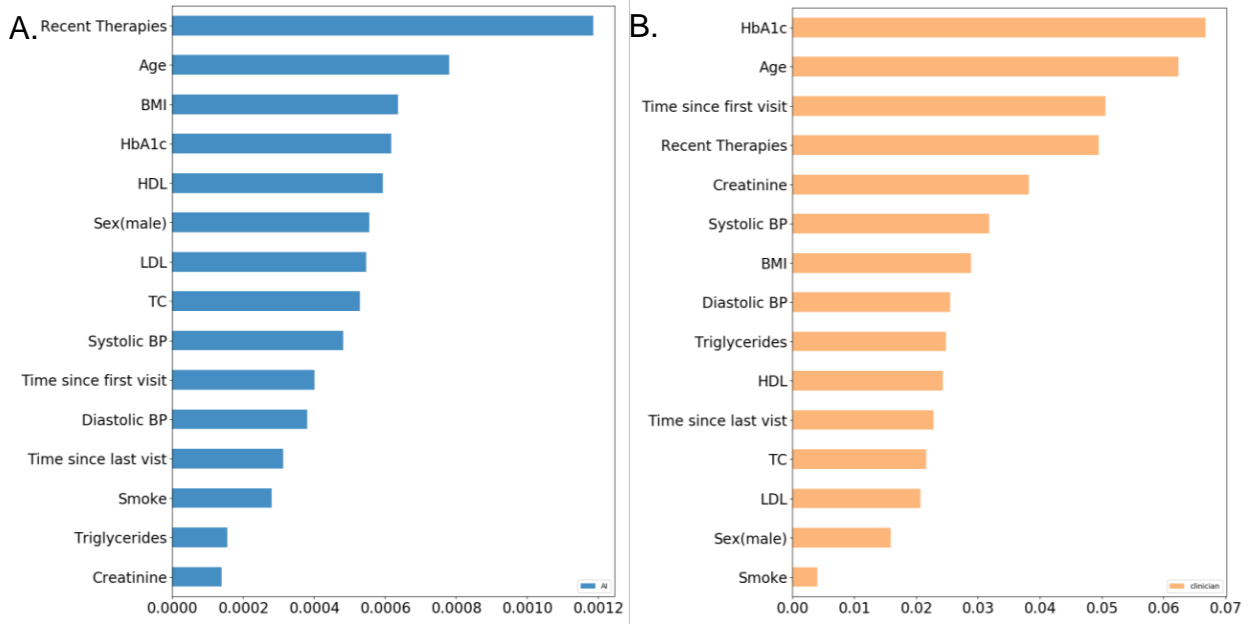

**Feature-importance of RL-multimorbidity (A) and clinician prescription (B).**

## 4. DISCUSSION

To our best knowledge, this is the first reinforcement learning assisted prescriptive algorithm for personalized single and multimorbidity outcome management for patients with T2DM. Using an EHR database, the developed RL algorithm can efficiently recommend treatment regimens to optimize patient health outcomes incorporating their individual demographic and treatment history. Compared with other machine-learning methods, the reinforcement learning approach has a particular advantage as it can efficiently learn complex dynamic drug-disease and drug-drug interactions in the presence of high temporal variation, uncertain outcomes and long-term treatment effects (15; 19). RL recommendations showed high levels of concordance with clinicians' prescriptions for single outcome optimizations of glycemia, blood pressure and CVD risk control. This demonstrates the feasibility of using RL for T2DM management and indicates that clinicians make near-optimal decisions with regard to single-outcome management.

RL-multimorbidity recommendations showed more frequent discrepancy with clinicians' prescriptions, as well as the recommendations by single-outcome RL algorithms. This provides data-driven evidence that optimizing multimorbidity management is different from optimizing single outcomes in parallel. For example, on the 1,272 patient encounters with the most frequently observed discrepancy between RL-multimorbidity and clinicians, their average A1c was 7.0%, SBP was 127.2 mmHg, and CVD risk was 12.6%. For these encounters, clinicians prescribed BIG to prioritize glycemic control while RL-multimorbidity prescribed HMG for lipid-lowering. This indicates challenges and uncertainties of multimorbidity management for patients with borderline and balanced levels of severities in multiple chronic conditions (26; 27). RL-multimorbidity showed overall improvements in managing the three outcomes simultaneously, significantly reducing the number of encounters with uncontrolled glycemia, uncontrolled HTN and high FRS CVD risk.

Although both clinicians and RL-multimorbidity place high importance on similar factors, these factors are differently ranked. RL algorithms did not weigh features that were not included in the reward functions, such as creatinine, as much as clinicians who consider it as an important renal function biomarker. This indicates a potential challenge of the RL algorithms using single-directed reward outcomes as the optimization goal. Ideally, a comprehensive reward function should incorporate domain knowledge and adverse events, such as hypoglycemia and kidney comorbidity, to achieve optimized outcomes while balancing risks of adverse events (28).

Typical limitations with EHR data are their unobserved medication adherence, partially observed clinical data at each encounter and uncontrolled time span between encounters (29). However, the RL algorithms were designed to incorporate these uncertainties under real-world scenarios. Particularly, if there were observable patient characteristics that associated with higher non-adherence to a certain treatment leading to lower levels of efficacy, RL would be able to identify it and prescribe different treatments for patients with those characteristics.

Although our evaluation methodology controls for several confounding factors that could explain differences in treatment effects, we can only estimate counterfactual outcomes under RL recommendations for patients with discrepant prescriptions. In addition, the T2DM patient population from NYULH ambulatory care may not be representative of the U.S. T2DM population. To ultimately validate the efficacy of the RL algorithms, randomized clinical trials with patients randomly assigned to RL and clinician mechanism would be needed.

## 5. CONCLUSIONS

In this study, we demonstrated feasibility of using reinforcement learning prescriptive algorithms for patients with type 2 diabetes mellitus to manage their multimorbidity based on test data from an ambulatory care center. The RL-glycemia, RL-BP and RL-CVD algorithms showed high concordance (83%-98%) with clinicians' prescriptions while RL-multimorbidity showed relatively low concordance (71%) for multimorbidity management. For patient encounters in which the RL recommendations differed from the clinician prescriptions, RL prescriptions showed significantly improved health outcomes as compared to clinicians' prescriptions. Potentially, the algorithm can be integrated into electronic health record platforms to assist physicians for T2DM management with dynamic real-time suggestions of personalized treatment paths.

**Title: Personalized Multimorbidity Management for Patients with Type 2 Diabetes Using Reinforcement Learning of Electronic Health Records**

**Supplemental Material**

**Reinforcement Learning Algorithms**

A Markov decision process (MDP) was used to model the decision-making process and approximate individual patient health trajectories. We formalize the MDP by the tuple $(\mathcal{S}, \mathcal{A}, P, r, \gamma)$, where

- $\mathcal{S}$ denotes a finite set of states, typically including patients' personal information, current/historical health status, current/historical treatment;
- $\mathcal{A}$ denotes the finite set of actions available given state $s$, including the set of regimens, where each regimen may include one or multiple therapies;
- $P$ represents the probability that taking action $a$ in state $s$ at time $t$ will lead to state $s'$ at time $t + 1$ (i.e., the patient's health state changes to $s'$ at $t + 1$ after taking regimen $a$ at time $t$), which describes the dynamics of the system;
- $r$ represents the immediate reward received for transitioning to state $s'$ (biomarker improvement);
- $\gamma$ denotes the discount factor, which makes immediate rewards more valuable than long-term rewards and determines the temporal impact of the current action: greater $\gamma$ indicates longer impact of current therapy action.

The process is observed at discrete time steps. In each time $t$, the agent observes the current state $s_t \in \mathcal{S}$ which typically includes the reward $r_t$. It then chooses an action from the set of available drugs $a_t \in \mathcal{A}$ and the patient health conditions moves to a new state $s_{t+1}$, and observes a feedback in form of a reward signal $r_{t+1}$ associated with the one-step transition $(s_t, a_t, s_{t+1})$. The medication selection rule is called the *policy* and denoted by a mapping $\pi$ from state space $\mathcal{S}$ to action space $\mathcal{A}$, i.e., $a_t = \pi(s_t)$. The quality of a policy is measured using the value function

$$V^\pi(s) = \mathrm{E}[\sum_{t=1}^{\infty} \gamma^t r_t | s_0, \pi] \quad (1)$$

which is defined as the expected cumulative discounted reward starting with state $s_0$, given that policy $\pi$ is used to make decisions. Then, the goal of a reinforcement learning agent is to learn the optimal policy $\pi^*$ which maximizes the expected cumulative discounted reward, that is, $\max_{\pi} V^{\pi}(s)$.

The reward $r_{t+1}(s_t, s_{t+1})$ at each time $t$ for each disease is defined as follows,

(1) For T2DM, given current state $s_t$ and medication $a_t$, we define the reward of $a_t$ based on $A1c_t \in s_t$ and $A1c_{t+1} \in s_{t+1}$ as

$$r_t^{T2DM} \equiv \varphi(s_t, s_{t+1}, \theta, \sigma) = \begin{cases} (A1c_{t+1} - A1c_t)\frac{A1c_t - \theta}{\sigma}, & A1c_t, A1c_{t+1} \geq \theta; \\ 0, & \text{otherwise} \end{cases} \quad (2)$$

where $\theta = 5.6$ is the diagnosis threshold for T2DM (1) and $\sigma = 1.58$ is the standard deviation of A1c in the cohort.

(2) For hypertension, given current state $s_t$ and medication $a_t$, we define the reward of $a_t$ based on $\text{SBP}_t \in s_t$ and $\text{SBP}_{t+1} \in s_{t+1}$ as

$$r_t^{HTN} \equiv \varphi(s_t, s_{t+1}, \theta, \sigma) = \begin{cases} (\text{SBP}_{t+1} - \text{SBP}_t)\frac{\text{SBP}_t - \theta}{\sigma}, & \text{SBP}_t, \text{SBP}_{t+1} \geq \theta; \\ 0, & \text{otherwise} \end{cases} \quad (3)$$

where $\theta = 120$ is the diagnosis threshold for hypertension and $\sigma = 17.7$ is the standard deviation of SBP in the cohort.

(3) For CVD, we use FRS global CVD function to predict total CVD outcomes (2) as indicator for CVD risk, denoted by $FRS_t$. The CVD risk factors considered in FRS include age, sex, race, ethnicity, smoking status and the following biomarkers: systolic BP (SBP), body mass index (BMI), A1c, total cholesterol (TC), and high-density lipoprotein (HDL). Smoking behavior (yes) was defined as either a previous smoker or current smoker. Then the reward function of taking $a_t$ is defined by the difference $r_t^{ASCVD} = FRS_t - FRS_{t+1}$.

(4) Multimorbidity. The reward for multimorbidity is defined by the average rewards from T2DM, hypertension and CVD, i.e.

$$r_t = \sum_{k \in \{\text{T2DM,HTN,CVD}\}} \frac{r_t^k - mean(r_t^k)}{sd(r_t^k)}, \quad (4)$$

where $mean(r_t^k)$ is the mean reward for $k$th disease and $sd(r_t^k)$ is the standard deviation of rewards for $k$th disease.

**Learning the Optimal Policy**

In reinforcement learning, many algorithms focus on estimating the so-called "Q-function" $Q^\pi(s, a)$ of a policy $\pi$ rather than the value function. The Q-function represents the expected value of state-action pairs, and can be connected to the value function through the equation

$$V^\pi(s) = \max_a Q^\pi(s, a), \tag{5}$$

that is, the optimal action is found by maximizing the Q-function. Thus, the Q-function measures the expected return or discounted sum of rewards obtained from state $s$ by taking action $a$ first and following policy $\pi$ thereafter. The *optimal* Q-function is then defined as the maximum return that can be obtained starting from observation $s$, taking action $a$ and following the optimal policy $\pi^*$ thereafter. The optimal Q-function is known to obey the following *Bellman* optimality equation:

$$Q^{\pi^*}(s, a) = \mathrm{E}_{s'}\left[r_t(s, s') + \gamma \max_{a'} Q^{\pi^*}(s', a') \,\middle|\, s, a\right]. \tag{6}$$

We use a nonlinear function, such as a neural network with parameters $\theta$, to approximate the action-value function, i.e., $Q^\pi(s, a) \approx Q^\pi(s, a; \theta)$. Such a neural network is called a Q-network (3). The approximation is trained by minimizing the difference (loss function) between the left- and right-hand side in Eq. (6), i.e.

$$L(s, a) = \mathrm{E}_{s' \sim p(\cdot|s, a)}\left[\left(Q^\pi(s, a; \theta) - r_t(s, s') - \gamma \max_{a'} Q^\pi(s', a'; \theta)\right)^2 \,\middle|\, s, a\right], \tag{7}$$

or equivalently,

$$L(s, a) = \mathrm{E}_{s' \sim p(\cdot|s, a)}[\ell_\theta(s, a, s')|s, a], \tag{8}$$

where

$$\ell_\theta(s, a, s') = \left(Q^\pi(s, a; \theta) - r_t(s, s') - \gamma \max_{a'} Q^\pi(s', a'; \theta)\right)^2$$

where $\text{target}(s') = r_t(s, s') + \gamma \max_{a'} Q^\pi(s', a'; \theta)$ is called the target value, $Q^\pi(s, a; \theta) - \text{target}(s')$ is called *TD error* and $p(\cdot|s, a)$ represents the state transition distribution. Ideally, we want the error to decrease, meaning that our current policy's outputs are becoming more similar to the true Q values.

Then, differentiating the loss function with respect to the weights with fixed target value we arrive at the following gradient

$$\nabla_{\theta_k} \ell_{\theta_k}(s, a, s') = \left(Q^\pi(s, a; \theta) - \text{target}(s')\right) \nabla_{\theta_k} Q^\pi(s, a; \theta). \tag{9}$$

When we update the target network, i.e., Q function $Q^\pi(s', a'; \tilde{\theta})$ in $\text{target}(s')$, every iteration makes learning computationally less stable. Therefore, the target value is fixed by using previous Q function parameter $\tilde{\theta}$, i.e., $\text{target}(s') = \text{target}(s'; \tilde{\theta}) = r_t(s, s') + \gamma \max_{a'} Q^\pi(s', a'; \tilde{\theta})$ and replacing the weight of the target network by the weight of the current Q-function $Q^\pi(s, a; \theta)$ every $C$ iterations, i.e., $\tilde{\theta} = \theta_k$.

The Q-network model developed in our paper uses a multi-layer feed-forward architecture which evaluates each state-action pair $(s, a)$. Specifically, the Q-network model contains three intermediate layers with 256 neurons in the first layer, 512 neurons in the second layer, 256 in the third layer, and one output layer containing varying numbers of neurons (equal to the number of actions) for RL-glycemia, RL-BP and RL-CVD and RL-multimorbidity. To prevent overfitting, we use the dropout method (4) as a regularization technique to selectively ignore single neurons during training. In the architecture, each dense layer is followed by a dropout layer with 0.5 dropout rate.

We also used the early stopping (5; 6) to prevent overfitting. The objective of DQN is to minimize the mean squared TD error, i.e., $\ell_\theta(s, a, s') = \left(Q^\pi(s, a; \theta) - r_t(s, s') - \gamma \max_{a'} Q^\pi(s', a'; \theta)\right)^2$. Another metric of interest is the consistency of recommendations between doctor and RL, quantified by the proportion of encounters that RL's recommendation agrees with clinicians' recommendations. In the study, we noticed that this second metric tended to converge 2500-5000 iterations later than the TD error. Thus, during training, we monitored both metrics and set the early stopping criterion to be that "mean squared TD error is not improved in last 5000 iterations".

Our training scheme is as follows:

1. We randomly selected 60% of the eligible patients as the training cohort to develop the RL algorithm and reserved the remaining 40% of patients as the test cohort to evaluate the performance of the RL algorithm.
2. We randomly selected 20% of the training data as a validation cohort. Then we trained deep Q network on the remaining 80% training data, stop training when its performance "mean squared TD error" was kept unimproved in 5000 consecutive iterations, and recorded the last iteration as L.
3. In the end, we obtained our final model by training the full Q network for L iterations.

In reinforcement learning, learning an optimal policy from observational data is referred as to batch mode RL (7). This approach uses a set of one-step transition tuples: $\mathcal{D} = \{(s_i, a_i, r_i, s_i'): i = 1, \ldots, |\mathcal{D}|\}$ to estimate the Q-function and thus obtain the prescription policy $\pi: a = \arg\max_{a'} Q^\pi(s, a'; \theta)$.

**Algorithm: batch DQN with random sampling**

Initialize training data $\mathcal{D}$. Initialize action-value function $Q^\pi(s, a; \theta)$ with random weights $\theta_0$. Initialize target action-value function $\hat{Q}^\pi(s, a; \theta)$ with random weights $\tilde{\theta}_0$. Set $C$

**for** $k = 1, 2, \ldots$ till convergence **do**

1. Randomly sample minibatch of $(s_i, a_i, r_i, s_i')$ from $\mathcal{D}$;
2. Set $\text{target}(s_i') = r_i + \gamma \max_{a'} Q^\pi(s_i', a'; \tilde{\theta})$;
3. Update $\theta$: $\theta_{k+1} \leftarrow \theta_k + \alpha_k \nabla_{\theta_k} \ell_\theta(s_t, a_t, s_{t+1}; \tilde{\theta})$ with Eq.(9) by applying the "Adam" stochastic gradient descent method;
4. Every $C$ iterations, set $\tilde{\theta} = \theta_k$.

**end**

**Data Imputation**

Our dataset contains a set of historically observed health states, but not every possible health state. In order to calculate an optimal policy, RL requires a way to estimate outcomes in any state, including those not in the original data. Following Lundberg et al.

(8), we imputed data for such states based on the information present in previous encounters using an exponentially decaying weighted average (EDWA). Since the chronic diseases and their treatments usually have lagged effect, and patient health status does not greatly change in a short period of time, we fill in the missing values using EDWA with a three-month moving window, 0.5 decay rate and 30 days half-life time. The decay rate specifies how much impact each past time point has on the computed mean for the time series.

**Evaluation**

The electronic health records (EHR) usually present a large number of variables. These variables frequently correlate with each other and have different scales. To address this problem, we use principal component analysis (PCA), an orthogonal transformation to represent sets of potentially correlated variables with principal components (PC) that are linearly uncorrelated. PCs are ordered so that the first PC has the largest possible variance and only some components are selected to represent the correlated variables; see the reference (9).

Then we apply $k$ nearest neighbor search (kNN) regression to predict the counterfactual health outcome for encounter denoted by $(s, a)$. For any given action $a$, we first query encounters in which the treatment regimen $a$ was prescribed by clinicians, then find $k$ nearest encounters (in Euclidean distance) to these in input encounter $s$, the predicted health outcome of an encounter is assigned by the mean of its neighbors' outcomes, i.e.

$$f(s) = \sum_{j \in N_k(s,a)} y_j, \qquad (10)$$

where $N_k(s, a)$ represent the $k$ nearest neighbors of state $s$ given action $a$, and $y_j$ denotes the biomarkers of interest of $j$th sample in validation set, such as A1c, SBP and FRS-CVD risk.

Specifically, given the validation data: $\mathcal{D}_v = \{(s_i, a_i, r_i, s'_i): i = 1, \dots, |\mathcal{D}_v|\}$, that is a set of one-step transition tuples that were not used for training, we first use PCA (9) to reduce the set of intercorrelated variables (i.e., state $s$) into a few dimensions accounting for 90% of the variance of state variables. These dimensions are called components and have the properties of collecting highly correlated variables within each component and being uncorrelated with each other. Let $c_i^k$ denote the $k$th principal component vector (PCs) for state $s_i$. Then the validation set of clinical encounters becomes $\mathcal{D}_v = \{(c_i, a_i, r_i, c'_i): i = 1, \dots, |\mathcal{D}_v|\}$, where $c_i = [c_i^1, \dots, c_i^k]$.

For encounter $i$, we applied our prescriptive algorithm to recommend a therapy, denoted by $a_i^{AI} = \pi^*(s_i)$. If that recommendation matched the prescribed standard of care therapy $a_i^{AI} = a_i$, we observed the true (historical) effect from the therapy. Otherwise, the outcome was imputed by k nearest neighbor search (kNN) regression by Eq. (10)., i.e., averaging the outcomes of the most similar patient encounters at which the recommended therapy $a_i^{AI}$ was administered.